\documentclass[12pt,preprint]{aastex}
\usepackage{amsmath}

\slugcomment{Not to appear in Nonlearned J., 45.}

\shorttitle{Multipole Lensing and High-order Perturbations}
\shortauthors{Chu et al.}

\begin{document}


\title{Multipole Gravitational Lensing and High-order Perturbations \\
       on the Quadrupole Lens}


\author{Z. Chu\altaffilmark{1,4}, W. P. Lin\altaffilmark{1,4}, G. L. Li\altaffilmark{2} and X. Kang\altaffilmark{3} }
\altaffiltext{1}{Key Laboratory for Research in Galaxies and Cosmology, Shanghai Astronomical Observatory,
Chinese Academy of Sciences, 80 Nandan Road, Shanghai 200030, China}
\altaffiltext{2}{Purple Mountain Observatory, 2 West Beijing Road, Nanjing 210008, China}
\altaffiltext{3}{The Partner Group of MPI for Astronomy, Purple Mountain Observatory, 2 West Beijing Road, Nanjing 210008, China}
\altaffiltext{4}{University of Chinese Academy of Sciences, 19A Yuquan Road, Beijing 100049, China}
\email{chuzhe,linwp@shao.ac.cn}

\begin{abstract}
An arbitrary surface mass density of gravitational lens can be decomposed into multipole components. We simulate the ray-tracing for the multipolar mass distribution of generalized SIS (Singular Isothermal Sphere) model, based on the deflection angles which are analytically calculated. The magnification patterns in the source plane are then derived from inverse shooting technique. As have been found, the caustics of odd mode lenses are composed of two overlapping layers for some lens models. When a point source traverses such kind of overlapping caustics, the image numbers change by $\pm 4$, rather than $\pm 2$. There are two kinds of images for the caustics. One is the critical curve and the other is the transition locus. It is found that the image number of the fold is exactly the average value of image numbers on two sides of the fold, while the image number of the cusp is equal to the smaller one. We also focus on the magnification patterns of the quadrupole ($m=2$) lenses under the perturbations of $m=3$, 4 and 5 mode components, and found that one, two, and three butterfly or swallowtail singularities can be produced respectively. With the increasing intensity of the high-order perturbations, the singularities grow up to bring sixfold image regions. If these perturbations are large enough to let two or three of the butterflies or swallowtails contact, eightfold or tenfold image regions can be produced as well. Possible astronomical applications are discussed.
\end{abstract}


\section{Introduction}

Strong gravitational lensing are the phenomena of gravitational light deflections that producing easily visible distortional images such as Einstein rings, arcs, and multiple images. The strong lensing allows us to probe the mass distribution of galaxies, groups and clusters, as well as invisible dark substructures \citep{dal02,rus03,yoo06,orb09}. Moreover, the magnification effect of the gravitational lensing also provides chances to study galaxies, black holes, and active nuclei that are too small or too faint to be resolved or detected with current instruments \citep{tre10}. One of the most distinct qualities of gravitational lensing is the image multiplicities of distant quasars or galaxies lensed by foreground galaxies or galaxy clusters. For non-singular lenses, it is well known that the total image number is odd \citep{bur81,sch92}. Because most of the dark matter halos are triaxial ellipsoid \citep{kas93,eva00,jin02}, their planar projections should correspond to the elliptic lenses \citep{kas93}. If a source lies within the central caustic of a typical elliptic lens (or quadrupole lens), there will be five images produced. However the image located near lens center is usually highly demagnified and faint \citep{rus01a,win04}, resulting in even image number of many observed quasars. \citet{kee00} and \citet{eva01} have demonstrated that sextuple or octuple images of lensed quasars are likely to exist, brought by the elliptic lens perturbed by external shears or the lens isophote deviated from pure ellipses such as boxiness or diskiness. B1359+154 is the first example of galaxy-scale gravitational lensing in which six images are observed of the same background quasar. The density configuration of the lens is complex which was proposed to include three primary lens galaxies \citep{mye99,rus01b}.

The critical curves and caustics are crucial for qualitatively understanding strong gravitational lensing. In general, the critical curves are closed curves in the lens plane (image plane), where the Jacobian matrices vanish and the magnifications are infinite. They divide the lens plane into image regions of positive and negative parities. The caustics are the corresponding curves obtained by mapping the critical curves into source plane via lens equation \citep{sch92,sch06}. Accordingly, the critical curves are the images of caustics. In fact, there are also other images of the caustics \citep{bla86,suy06} and they are the so-called transition loci \citep{fin02} which receives less attention so far. Nevertheless, they are important for us to understand the spatial distribution of multiple images of strong lensing, as is one of the main topics of this paper. The caustics of typical elliptic lens can be divided into radial and tangential caustics. The tangential caustic commonly comprises cusp points and fold lines, where the cusps are the generic singularities. In addition to the cusp, there are some other high-order singularities, such as swallowtail and butterfly. The detailed information about the singularity theories in strong gravitational lensing can be obtained elsewhere (e.g. \citealt{pet93,aaz09,orb09}). The caustics divide the source plane into regions with different image multiplicities. When a source traverses a single caustic, either fold or cusp, the total number of images changes by $\pm$2. However, in the case that the caustics are composed by two overlapped layers, the situation will be different as presented later in the main text of this paper.

In practice, decomposing the projected mass distribution or planar potential into multipole modes is an often used method to study gravitational lens (e.g. \citealt{tro00,eva03,koc04,ber09}). In addition to the monopole ($m=0$), the quadrupole ($m=2$) is usually the most important component and has been investigated intensively \citep{kov87,kor94,wit00,fin02}. The components with order higher than 2 are almost not dominant in the expansions of real lenses, and are usually more complex than the quadrupole. Nevertheless, their perturbation may have significant impacts on image properties and should not be neglected totally. For this purpose, by using analytical and numerical methods, we focus on the properties of these high-order components, and then study their perturbations on the $m=2$ lenses with different intensities and phase differences. Although it is not clear if there are such strong high-order perturbations in reality, it is still worth of investigations for theoretical foreseeing of multiple image lenses.

The paper is arranged as follows. In Section 2, we investigate the properties of the $m=2$, 3, 4 and 5 mode gravitational lenses. Different image multiplicities and overlapped caustics are investigated. In Section 3, the images of the caustics including critical curves and transition loci are studied. Then we discuss how the image number changes when a point source traversing the overlapped caustics. In Section 4, we focus on the magnification patterns of the quadrupole ($m=2$) lenses perturbed by the $m=3$, 4 and 5 mode components. Then we interpret how the swallowtail and butterfly singularities appear and evolve under the perturbations with different intensities and phase differences. Finally in Section 5, conclusions of the paper and a discussion are given.

\section{Properties of the Multipole Lensing}

\subsection{Multipole Decomposing and Lens Equations}

Under the condition that the size of lens object is much smaller than the distances from the lens to observer and source, the lens can be generally assumed as a thin mass plane. A very useful way to describe lensing mass distribution is multipole expansion. Through this method, an arbitrary mass surface density or convergence $\kappa(\boldsymbol{\theta})$ can be decomposed into multipole components \citep{hut05,sch06} with following function
\begin{equation}
\kappa(\boldsymbol{\theta})=\kappa_{0}(\theta)+\sum_{m=1}^{\infty}\{\kappa_{m}(\theta)\cos m[\phi-\phi_{m}(\theta)]\} .
\end{equation}
Here $(\theta, \phi)$ are the polar coordinates, and $\phi_m(\theta)$ is the phase for the mode $m$ at radius $\theta$. \begin{equation}
\kappa_{0}(\theta)=\frac{1}{2\pi}\int_0^{2\pi} \kappa(\boldsymbol{\theta})d \phi
\end{equation}
is the monopole ($m=0$) component of the surface mass density.

To simplify the multipole lensing problem, we apply the familiar Singular Isothermal Sphere (SIS) lens model as the monopole ($m=0$) distribution. Note that this spherical mass distribution yields flat rotation curves \citep{sch92,mao93}. Then one multipole component with a radial profile similar to the SIS is added to the $m=0$ mode. As first step, for simplicity, the phase of the multipole $\phi_{m}(\theta)$ is set to 0. Therefore, the multipolar mass surface density of generalized SIS model is given by
\begin{equation}
\kappa(\boldsymbol{\theta})=\frac{\theta_{E}}{2\theta}(1+k_{m}\cos m\phi) ,
\end{equation}
where the $\theta_{E}$ is the Einstein radius of the SIS lens model. Hence, $\kappa_0(\theta)=\theta_E/2\theta$ represents the profile of $m=0$ mode component, and $\kappa_m(\theta)=k_m\kappa_0(\theta)$ ($0\leq k_m\leq1$) expresses the radial profiles of the multipolar components.

Before ray-tracing simulations with the surface density field $\kappa(\boldsymbol{\theta})$, one must know the deflection (or lensing) potential and the deflection angle as a function of position in the lens plane. The deflection potential depends on the two dimensional Poisson equation $\nabla^2\psi(\boldsymbol{\theta})=2\kappa(\boldsymbol{\theta})$. By solving this differential equation in polar coordinate, we can derive the deflection potential
\begin{equation}
\psi(\boldsymbol{\theta})=\theta_{E}\theta(1-\frac{k_{m}}{m^2-1}\cos m\phi) , m\neq1 ,
\end{equation}
\begin{equation}
\psi(\boldsymbol{\theta})=\theta_{E}\theta(1+\frac{k_1}{2}\ln\theta\cos\phi) , m=1 .
\end{equation}
Hence, the scaled deflection angle $\boldsymbol{\alpha}=\nabla\psi(\boldsymbol{\theta})$ is the first derivation of the deflection potential. It can be written as radial and tangential components
\begin{equation}
\alpha_{rad}=\theta_{E}-\frac{\theta_{E}k_{m}}{m^2-1}\cos m\phi , m\neq1 ,
\end{equation}
\begin{equation}
\alpha_{tan}=\frac{m\theta_{E}k_{m}}{m^2-1}\sin m\phi , m\neq1 ,
\end{equation}
\begin{equation}
\alpha_{rad}=\theta_{E}+\frac{\theta_{E}k_1}{2}\cos \phi+\frac{\theta_{E}k_1}{2}\ln\theta\cos\phi , m=1 ,
\end{equation}
\begin{equation}
\alpha_{tan}=-\frac{\theta_{E}k_1}{2}\ln\theta\sin\phi , m=1.
\end{equation}
Based on the deflection potential $\psi$, we found the shear $\gamma=\kappa$ for the modes $m\neq1$ (using Equation [13] in the paper \citealt{ber09}). Consequently, the magnification in the lens plane is
\begin{equation}
\mu=\frac{1}{(1-\kappa)^2-\gamma^2}=\frac{1}{1-2\kappa} ,
\end{equation}
and therefore the critical curves where $\mu$ is infinite is given by
\begin{equation}
\theta=\theta_{E}+\theta_{E}k_{m}\cos m\phi .
\end{equation}

Alternatively, the deflection potential $\psi$, the deflection angle $\boldsymbol{\alpha}$ and the shear $\gamma$ can also be calculated numerically by Fast Fourier Transform method via $\kappa(\boldsymbol{\theta})$ (refer to \citealt{bar10} for more details). Solving these equations in Fourier space imposes periodic boundary conditions. The lens grids could be set in the center of a larger grids, while the densities outside the inner grid are padded with zeros. The larger grid must be at least twice as large as the inner grid, and the larger the better \citep{li05,ama06}. In this work, the inner grid size of the ray-tracing simulation is $1024\times1024$, corresponding to the resolution about $0.004\theta_{E}$, which could produce results sufficiently consistent with those derived from the analytical method.

Finally, the lens equation $\boldsymbol{\beta}=\boldsymbol{\theta}-\boldsymbol{\alpha}$ describing the transformation between the lens plane $(\theta, \phi)$ and the source plane $(\beta, \varphi)$ can also be written in the polar coordinate as
\begin{equation}
\beta^2=(\theta-\alpha_{rad})^2+\alpha_{tan}^2 ,
\end{equation}
\begin{equation}
\tan(\phi-\varphi)=\frac{\alpha_{tan}}{\theta-\alpha_{rad}} .
\end{equation}

\subsection{Image Multiplicities of the Multipole Lenses}

We simulate the multipole lenses of modes $m=2$, 3, 4 and 5, corresponding to the quadrupole, sextupole, octupole and decagonal pole modes respectively. The monopole lens ($m=0$) is the circular symmetric lens, and this simple lens model has been well known to us. A dipole moment ($m=1$) corresponds to making the lens lopsided with more mass on one side of the center than the other, or equivalently, a shift in the center of mass. The dipole moment vanishes if the coordinate origin is shifted to this point \citep{dom99,sch06,jog09}. Furthermore, the expressions of the deflection angle for $m=1$ is not similar to the other modes in appearance, and it is divergent when the $\theta$ tends to infinity in Equations (8)-(9). Therefore, only the $m=2$, 3, 4 and 5 modes of lens components were taken into account through out the paper.

Magnification patterns can well exhibit the caustic profiles in the source plane, and be computed by the inverse ray-shooting technique \citep{kay86,sch87,med06}. The main ideal is shooting a bundle of light rays from observer to each grid of the lens, then the light deflecting to the source plane according to deflection angle $\boldsymbol{\alpha}$. The place in source plane where irradiated by more light rays corresponds to larger area of the lens plane, and it means higher magnification levels there. Accordingly, these patterns present the tangential caustics of the multipole lenses clearly. Because a singular isothermal lens does not formally have a radial critical curve \citep{eva98,hut05}, the radial caustic (or pseudo-caustic in this situation) associated with the center singularity in lens plane will not be discussed. We only focus on the tangential caustics and assume there is no naked cusp. After deriving the magnification patterns, we set them as source maps, so their images can be obtained through ray-tracing simulations (i.e., mapping these magnification patterns to the lens plane). The magnification patterns and their images are shown on the left and middle columns of Figure 1. The right column displays the magnification maps in the lens plane which are calculated through Equation (10).

The magnification patterns of the $m=2$, 3, 4 and 5 modes are shown in the left column from top to bottom respectively, where $k_m$ is set to 0.3 in each case. The numbers on the magnification patterns denote different multifold image regions, and the numbers on the middle column denote the different image regions with reference to the left column. Due to the extraordinary complexities brought about by numerous intersecting folds in the caustics of the $m=4$ and 5 mode lenses, their corresponding multifold image regions are not denoted in Figure 1 (h) and 1 (k). Since the middle column shows the images of the magnification patterns, they can well present the critical curves and transition loci, which are the images of the caustics \citep{fin02}. The magnification maps on the right column can only show the critical curves by high magnification red regions, not including the information about transition loci.

In Figure 1 (a), one can see the familiar astroid caustic, inside of which is 4-image region, while outside of which is 2-image region (hereafter, the 1-image region outside of the pseudo-caustic is not considered). The cusps can be divided into major cusps and minor cusps. A source near the major cusp can be mapped into three close images around the tangential critical curve on the same side of the source with respect to the lens center, while a source on the minor cusp can be mapped into three close images on the opposite side of the source \citep{xu09}. In Figure 1 (a), the left and right cusps are the major cusps, while the top and bottom ones are the minor cusps. The areas near major cusps are slightly redder than those of minor cusps, which means that the formers have higher level of magnifications. Figure 1 (g) shows the tangential caustic composing eight cusps. The four cusps in the horizontal and vertical directions are the major cusps, while the four cusps in the $\pm45^\circ$ directions are the minor cusps. As the numbers denoted in Figure 1 (g), it can yield eight, six, four and two images in different regions from inner to outer side of the magnification pattern. On the other hand, as shown in Figure 1 (d) and 1 (j), for the magnification patterns of $m=3$ and $m=5$ modes, the tangential caustics only show three and five cusps respectively, instead of six and ten. In these two cases, the differences of the image numbers between two sides of the caustics are 4.

In fact, the cusp numbers for the $m=3$ and 5 are reduced by half because the major and minor cusps overlap with each other. If we put the Equations (6)-(7) and (11) into Equations (12)-(13), the expression for the caustics in the form of parametric equations can be derived as
\begin{equation}
\beta=\frac{m\theta_{E}k_{m}}{m^2-1}\sqrt{m^2\cos^2 m\phi+\sin^2 m\phi} ,
\end{equation}
\begin{equation}
\varphi=\phi-\arctan\frac{\tan m\phi}{m}-(\pi) .
\end{equation}
Because the arc tangent function has a phase uncertainty for the shift of $\pi$, the $\pi$ in the bracket only needed when the $\theta<\alpha_{rad}$ in Equation (13), i.e., $\cos m\phi<0$ in Equations (6) and (11). Coupled with Equation (15), one can find that $\varphi(\phi+\pi)=\varphi(\phi)+\pi$ if $m$ is even, and $\varphi(\phi+\pi)=\varphi(\phi)$ if $m$ is odd. It is also easily obtained that $\beta(\phi)=\beta(\phi+\pi)$ from Equation (14). Thus, for the odd mode lens, the two points $\theta(\phi)$ and $\theta(\phi+\pi)$ on the critical curves correspond to the same point on the caustics, which means these caustics are overlapped of two layers. Furthermore, the major (minor) cusps correspond to the points $\cos m\phi=\pm1$ on the critical curves, so the distances from both major and minor cusps to the center of the source plane are exactly
\begin{equation}
\beta_{cusp}=\frac{m^2\theta_{E}k_{m}}{m^2-1} .
\end{equation}
Therefore, in the left column of Figure 1, the major and the minor cusps show the same distances to the source center.

Figure 2 is similar to Figure 1, except that we set $k_m=0.6$, which is as twice as the value used in Figure 1. From top to bottom, the left column lays out the magnification patterns for modes $m=2$, 3, 4 and 5 respectively, while their images are shown in the middle column. There are white squares in each center of the magnification patterns, located in 4-, 6-, 8-, and 10-image regions in turn. The red squares are located in the 2-, 2-, 4-, and 6-image regions in turn. As can be seen, in the middle column, there are four, six, eight, ten white images, and two, two, four, six red images from top to bottom respectively. The white images lie approximately on a circle, and in other words, they appear like incomplete Einstein rings. Additionally, the white areas outside the critical curves locate slight farther from the lens centers than those inside. The red square in Figure 2 (g) locates near the minor cusp, so that its three close images are on the opposite side of the source. The red square in Figure 2 (j) locates near the overlapping cusps (a major and a minor), so there are three close images on the same side of the source and the other three on the opposite side. Since the multipole intensity used in Figure 1 ($k_m=0.3$) is smaller than that used in Figure 2 ($k_m=0.6$), the critical curves of the former are rounder than those of the latter. If the intensities of the multipoles meet the condition of $k_m=0$, the critical curves will turn into Einstein rings. In addition, because the $\beta_{cusp}$ in Figure 2 are twice as large as those in Figure 1 (they are not drawn according to the proportion in the two Figures), the areas or the probabilities to yield more than two images of a single source for the cases in Figure 2 would be much larger than those in Figure 1.

\section{The Images of the Caustics}

Caustics are the divisions of different image multiplicities in the source plane, and can be mapped into lens plane to get their images. Critical curve is the most familiar one, but there are other types as introduced in the first section. \citet{fin02} have studied the images of the caustics for the $m=2$ lens by analytical method. We extend further by using numerical methods. Figure 3 shows the caustics of the $m=2$ and 3 mode components as well as their images. Due to many intersecting folds in the caustics of the $m=4$ and 5 mode lenses, we will not study them intensively. The investigations on the images of the caustics can be divided into two parts, one of which is for the folds and the other is for the cusps.


As shown in Figure 3 (a) and (b), each fold of the $m=2$ lens has three images. The images of different folds are distinguished by four different colors. Except the solid thick critical curve, there is dashed/dotted curve outside/inside of the critical curve. They are outer and inner 2-4 (2-image region to 4-image region, similar hereafter) transition loci, which can also be found in Figure 1 (b) and Figure 2 (b). In Figure 3 (c) and (d), one can find each fold in the $m=3$ mode lens has four images. As a result of the overlapping caustics, there are two critical curves, as indicated by solid thick lines in Figure 3 (d). The one indicated by dashed curve is outer 2-6 transition locus, and the one indicated by dotted curve is inner 2-6 transition locus. They can also be found in Figure 1 (e) and Figure 2 (e). Therefore, we can take the folds of the $m=2$ lens as 3-image regions, which divide the lens plane into 2-image and 4-image regions, and the folds of the $m=3$ lens as 4-image regions, which divide the lens plane into 2-image and 6-image regions. It is found that the image number of the fold is the average value of the two sides of the fold. Among these images, the number of the critical curves is the layer number of the fold, and the rest images of the fold are the transition loci in the lens plane.

It is intriguing when a source traverses such overlapped fold. For example, in Figure 2 (j), there are red and white squares in the 6-image and 10-image regions respectively, and they have six and ten images as shown in Figure 2 (k) correspondingly. The fold between the two squares has eight images including two critical curves and six transition loci. If this red square traverses the overlapping fold into the 10-image region, the left three images will move to the left traversing the three 6-10 transition loci into the 10-image regions, as the right three ones move to the left into the 10-image regions simultaneously. Two new images will appear on the critical curves denoted by two white arrows, and then each of them split into two images, resulting in four new images once the source is well within the 10-image region. At last, the total images number is ten. The process is similar for reverse traversing but with image eliminating on the two critical curves.


The cusps of the caustics in Figure 3 can be distinguished by the crossover points of different color folds. As shown in Figure 3 (b), each cusp of $m=2$ lens has two images, one of which is the tangent point of the critical curve and a transition locus, and another of which is the cusp of the other transition locus. Each cusp of $m=3$ lens also has two images. As shown in Figure 3 (d), one image is the tangent point of the critical curve and the outer transition locus, i.e., the image of the major cusp, locating on the same side of the source. The other image is the tangent point of the critical curve and the inner transition locus, i.e., the image of the minor cusp, locating on the opposite side of the source. Therefore, we can regard the cusps of the modes $m=2$ and $m=3$ as 2-image regions. It was found that the image number of the cusp is equal to the smaller of the image numbers on either side. Similarly to the folds, there are also two kinds of images. One is the tangent point of the critical curve and transition locus, and their image number is the layer number of the cusp. The outer and inner tangent points relative to the critical curve correspond to the major and minor cusps respectively. Another kind of images are the cusps of the transition loci.

In Figure 2 (j), the cusp located on the right side of the red square has two images, because it must be consistent with the image number of 2-image region. If the red square in the 6-image region moves to the right to traverse the overlapped cusp into the 2-image region, as implied in Figure 2 (k), the left three images will move to the right and merge into one image on the tangent point of the critical curve and transition locus, while the right three images will also merge into one image on the right tangent point simultaneously. Then the two merged images will move into the 2-image regions respectively. At last, there will be only two images in total. The process is similar for reverse traversing, but with each of the two images splitting into three images.

\section{Quadrupole Lenses Perturbed by High-order Modes}

In the expansion of projected real galaxies or simulated dark matter halos, besides the $m=0$ component, the $m=2$ component is usually the most important one. Although the intensities of $m=3$, 4 and 5 mode components hardly exceed the $m=2$ mode, their influences on the strong gravitational lensing are still significant and can be directly reflected in the perturbations on the caustic of the $m=2$ lens. For these higher-order modes, a non-zero phase $\phi_m$ must be included since circular symmetry is broken by the $m=2$ mode. In reality, the intensities and the phases of these high-orders are all functions of the radius. However, to simplify the perturbation problem, we assume that $\phi_m$ do not depend on the radius. In these cases, the matter distribution formula of $m=2$ plus high-order mode ($m=3$, 4 and 5) is given simply by
\begin{equation}
\kappa(\boldsymbol{\theta})=\frac{\theta_E}{2\theta}[1+k_2\cos 2\phi +k_m\cos m(\phi-\phi_m)] ,
\end{equation}
and the deflection potential is given by
\begin{equation}
\psi(\boldsymbol{\theta})=\theta_{E}\theta[1-\frac{k_2}{2^2-1}\cos 2\phi-\frac{k_{m}}{m^2-1}\cos m(\phi-\phi_m)] .
\end{equation}
Hence, the scaled deflection angle is also derived
\begin{equation}
\alpha_{rad}=\theta_{E}-\frac{\theta_{E}k_2}{2^2-1}\cos 2\phi-\frac{\theta_{E}k_{m}}{m^2-1}\cos m(\phi-\phi_m) ,
\end{equation}
\begin{equation}
\alpha_{tan}=\frac{2\theta_{E}k_2}{2^2-1}\sin 2\phi+\frac{m\theta_{E}k_{m}}{m^2-1}\sin m(\phi-\phi_m) ,
\end{equation}
and the critical curve is
\begin{equation}
\theta=\theta_{E}+\theta_{E}k_{2}\cos 2\phi+\theta_{E}k_{m}\cos m(\phi-\phi_m) .
\end{equation}

We can put the Equations (19)-(21) into Equations (12)-(13) to get the caustics, but it is not easy work because the expression is very cumbersome. However, it is very convenient to derive the magnification patterns by the inverse shooting technique through the deflection angle and lens equation. The results are shown in Figure 4, 5 and 6 for the modes $m=3$, 4 and 5 respectively. We then discuss the three kinds of perturbations respectively, based on the magnification patterns.

The $m=3$ component sometimes plays a very important role through its perturbation on the $m=2$ mode (e.g. \citealt{irw06}). We examine the perturbations of $m=3$ mode with different relative intensities $k_3$ and phases $\phi_3$. In Figure 4, the three rows from top to bottom correspond to the anticlockwise rotations of the $m=3$ component for the phases $\phi_3$ of 0, $\pi/4$, $\pi/2$ respectively. The sum of $k_2$ and $k_3$ is kept constant at 0.5, while $k_3$ increases from 0.0 to 0.5 from left to right. As shown on the magnification patterns of the top row in Figure 4, the left major cusp begin to shrink under a small perturbation, and a butterfly singularity appears. Then, the top and bottom two minor cusps become tilted to the left. The butterfly structure becomes more obvious with stronger perturbation, and the triangular pattern grown from the butterfly moves to the primary caustic and overlap with it eventually. It also confirms that the $m=3$ caustic is composed by two overlapped layers in triangular shape. The middle row of Figure 4 shows the patterns of $m=3$ mode rotated $\pi/4$ anticlockwise. Instead of a butterfly catastrophe, a swallowtail catastrophe appears on the fold. With increasing $m=3$ component, the triangle evolved from the swallowtail also moves and finally overlaps with the primary caustic. The bottom row of Figure 4 shows the results with a phase of $\pi/2$. In this case, the changing of patterns is similar to the top row, but with a butterfly evolved from a minor cusp. The $m=3$ perturbation can produce six images via a butterfly or swallowtail catastrophe. In general, the area of the sixfold image region grows with the increasing of the perturbation intensity.

The $m=4$ component is relatively more universal in strong lenses, such as the elliptical galaxy models with boxy or disky isophotes \citep{naa99,eva01}. Given that the $m=4$ mode does not change after rotating $\pi/2$, the three rows from top to bottom present the results with phase shift of 0, $\pi/8$, $\pi/4$ respectively. As before, the sum of $k_2$ and $k_4$ is kept constant at 0.5, while $k_4$ increases from 0.0 to 0.5 from left to right. As shown on the magnification patterns of the top row in Figure 5, the butterflies appear at top and bottom minor cusps. The two butterflies become more obvious and begin to contact with each other with increasing perturbation. At last, the magnification pattern gradually evolves to have eight distinct cusps. The middle row of Figure 5 shows the case with a phase of $\pi/8$. There are no butterflies, but swallowtails. With the increasing of $m=4$ component, the magnification pattern also evolves into the pure $m=4$ caustic. The bottom row shows the results of the $m=4$ perturbation rotated $\pi/4$ anticlockwise. They are similar to the top row but the two butterflies are evolved from two major cusps. The $m=4$ perturbation can also produce six images by the butterflies or swallowtails. If the perturbation is large enough to let the butterflies or swallowtails contact, it can produce eightfold image regions. In general, the areas of the sixfold or eightfold image regions also grow with the increasing of perturbation intensity. \citet{wit00} and \citet{kee00} have derived similar results, using analytical method to study the elliptic lens perturbed by an external shear with different angles and intensities. \citet{eva01} also carried out similar researches by studying boxy and disky isophotes deviated from the pure ellipses. Our results for perturbations with phase of 0, $\pi/8$, $\pi/4$ are equivalent to theirs for diskiness, skewed squarishness, and boxiness respectively.

The $m=5$ mode component is rarely investigated in strong gravitational lensing. Nevertheless, we study it in the way similar to $m=3$ and 4 modes. The three rows from top to bottom show results with anticlockwise rotations of 0, $\pi/4$, $\pi/2$ respectively. The components of $k_2$ and $k_5$ were still set as before. As shown in the top row of Figure 6, a butterfly appears at the left major cusp, and two swallowtails appear on the right folds. With increasing of the perturbation, the three structures begin to contact with each other. At last, the magnification pattern gradually evolves to the pattern of $m=5$. The middle row in Figure 6 shows that there are three swallowtails but no butterfly. With increasing perturbation, the magnification pattern also evolves to the pure $m=5$ mode caustic. The bottom row shows the result similar to that of the top row, but the butterfly is evolved from a minor cusp. The perturbation of the $m=5$ mode can also produce sixfold image regions by the butterflies or swallowtails. If the perturbation is large enough to let two of the butterflies or swallowtails contact, it can produce eightfold image regions. Similarly, if three of the butterflies or swallowtails contact, tenfold image regions are produced. In general, the areas of the sixfold, eightfold or tenfold image regions grow with the increasing of perturbation intensity. However, in this ideal case, eightfold image region disappears at last.

Butterfly and swallowtail are high-order singularities on the caustics. The swallowtail structure evolves on a fold, and it can increase two cusps. The butterfly structure was evolved from a cusp and has three cusps, so it also corresponds to increasing two cusps \citep{orb09}. The numbers of the cusps for the modes $m=2$, 3, 4 and 5 are four, six, eight and ten respectively (see Figures 1 and 2, and the overlapped cusp has been counted as two). Thus, for the $m=2$ mode lens perturbed by $m=3$, 4 and 5 modes, the number of the cusps mutate from four to six, eight and ten respectively. Therefore, one, two and three swallowtail or butterfly singularities should be yielded under such perturbations, as has been shown in Figures 4, 5 and 6. Additionally, since the numbers of the major and the minor cusps are equal in each mode of the lenses (see Figure 1 and 2), each pair of the increased cusps by a butterfly or a swallowtail must be one major and the other minor. The number of butterflies or swallowtails as well as their positions are determined by the phase differences between the high-order modes and the $m=2$ mode.

\section{Conclusions and Discussion}

The aim of this paper is to examine the properties of the pure multipoles of strong gravitational lenses and high-order perturbations on the $m=2$ lens. At first, we assume a simplified multipole surface distribution of mass $\kappa$, and then calculate the deflection potential $\psi$ and deflection angle $\boldsymbol{\alpha}$ using analytical and numerical methods. The ray-tracing simulations based on the two ways give consistent results, but there is no doubt that using the analytical deflection angle should be more accurate and time-saving. Through the deflection angle and lens equation, the magnification patterns of the source plane are derived by inverse shooting technique. These patterns display the tangential caustics in the source plane clearly. Therefore, the tangential caustics of these multipole lenses, and their corresponding images including transition loci in the lens plane can be studied in great details. It is found that the lenses of the $m=2$, 3, 4 and 5 modes, with singular points in the lens center, can produce four, six, eight and ten images of a single source at most respectively.

The quadrupole mass distribution could produce 4-image regions, and the generalized quadrupole includes the $m=2$ lens, the elliptic lens, and the binary lens. The sextupole distribution could produce 6-image regions, and similarly, the generalized sextupole includes the $m=3$ lens, and the triangular distribution of three galaxies. The lens objects of B1608+656 are two galaxies, which provide a quadrupole component and generate four images of a single source \citep{sur03,suy06}. The quasar B1359+154 at $z_s=3.235$ has six images, and they are yielded by three lens galaxies at $z_l\simeq1$ \citep{rus01b}. This example is not the same type with the $m=3$ lens in this work. However it gives strong support to our sextupole lens model. In Figure 1, we can find even for the same $k_m$, the multifold image region with the largest image number decreases with increasing $m$. Thus, the regions of image number larger than ten could be produced similarly, but the possibilities become lower and lower. Although these multipole lens models are very simple here, they provide new opportunities to study the image multiplicities of strong gravitational lensing, which could hardly be acquainted by general quadrupole lenses.

The expression of magnification $\mu$ for the multipole lenses are calculated. In addition, the critical curves and the caustics are also derived analytically. Through these analytical results of the generalized SIS lens model, we studied the overlapped caustics and their images. The caustics of odd mode lenses consist of two overlapping layers as shown in the simulations, and these overlapping phenomena are also theoretically proved. If a point source traverses such kind of caustics, the image number change by $\pm 4$, rather than $\pm 2$. The overlapped caustics could not be detached by the odd mode perturbations, but can be detached easily by even modes. Since the possible internal or external perturbations are very complex, the overlapped caustics are indubitably not universal in the real cases of gravitational lensing.

The images of the caustic include not only the critical curves but also the transition loci \citep{fin02}. The critical curves divide the lens plane into regions of positive and negative parities, while the transition loci divide the lens plane into regions of different image multiplicities. In general, the tangential caustic commonly comprises cusp points and fold lines. The image number of the fold is the average of the numbers on both sides of the fold, among which the number of critical curves is the layer number of the fold. The rest images are the transition loci in the lens plane. The image number of the cusp is equal to the smaller of the image numbers on either side of the cusp. Generally speaking, there are also two ways for the images of cusp to appear. One is the tangent point of the critical curve and transition locus, and the number is the layer number of the cusp. Another category is the cusp of transition locus. The tangent point of critical curve and transition locus can also be interpreted to be two images but contacted together. This is the reason that the image number of the cusp is smaller than that of the fold. The outer tangent point relative to the critical curve corresponds to the major cusp, while the inner tangent point corresponds to the minor cusp.

For the planar projections of the dark matter halos, besides the $m=0$ component, the $m=2$ mode is usually the largest one. In reality, there is almost no lens dominant by $m=3$, 4 and 5 modes, while their effects can be embodied in the perturbations on the $m=2$ lens. In this paper, we study the magnification patterns of the mode $m=2$ with perturbations by the $m=3$, 4 and 5 components. One, two and three swallowtail or butterfly singularities are yielded under these high-order perturbations. Additionally, each pair of increased cusps by a butterfly or a swallowtail must be one major and the other minor. The number for the butterflies or the swallowtails as well as their positions are determined by the phase differences between the high-order modes and $m=2$ mode. With the increasing of high-order perturbations, the butterfly or swallowtail singularities grow up to bring sixfold image regions. If these perturbations are large enough to let two or three of the butterflies or swallowtails contact, eightfold or tenfold image regions will appear. In general, the stronger the high-order perturbations, the larger areas of the multifold image regions. These results are consistent with and also the complements of the sextuplet and octuplet image theories derived by \citet{kee00} and \citet{eva01}.

Studying the multipole lenses could also help us to understand the number of giant arcs produced by sources locating near the caustics. Because high-order mode lenses can produce caustics with longer circumference and more cusps, the probabilities to generate giant arcs is higher than that of low-order lens. Furthermore, even a small high-order perturbation on the quadrupole lens could also increase the length of caustics and the number of cusps, by bringing butterfly or swallowtail structures. Therefore, we conclude that dark matter halos which are more complex and have stronger multipole components should have higher probability to generate giant arcs than the simple quadrupole lens. In deed, the quadrupole lens can also yield more arcs than the monopole lens (i.e., circular lens) for the same reason, which has been illustrated by \citet{men07}, who found that the elliptical lens with realistic density profiles produce a number of arcs larger by a factor of ten than circular lenses with the same mass. Substructures in dark matter halos could play vital roles in generating high-order singularities and increasing the probability to produce giant arcs \citep{li06,men07}. These effects could also be interpreted to be brought by the multipole components introduced by large substructures. Additionally, real lens should possibly include external shears produced by neighbor galaxies or merging clusters. External shears on the lens system can stretch the shape of the caustics, and can also add swallowtail or butterfly catastrophes \citep{cha84,sch86,kee00}. Therefore, the external shear should also be able to increase the probability to generate arcs.

The cases discussed in this work are all simplified exercises. In these models, the strengths of the multipoles fall off with the radius as SIS lens and phase dependence on radius are neglected. Actually in the real situation, the multipole components with many other radial profiles and phase dependence could also generate high-order singularities or multifold image regions. Nevertheless, they may not be resolved analytically and may not provide the situation of overlapped caustics. Realistic gravitational lenses can be decomposed into many multipole components, which have complex radial profiles and phase dependence on the radius. Thus, more complex and accurate modeling should take into account these realistic conditions in future investigations. The magnification patterns and the images of them for any arbitrary dark matter halos can be derived from ray-tracing simulations. Therefore, some useful information such as multifold image regions could be presented in the images of magnification patterns, so it may provide a powerful tool in theoretic researches of the multiplicities of strong gravitational lensing.



\acknowledgments

The authors are grateful to the anonymous referee for constructive comments and detailed suggestions to improve this manuscript. WPL acknowledges supports by the NSFC projects (No. 10873027, 11121062) and the Knowledge Innovation Program of the Chinese Academy of Sciences (grant KJCX2-YW-T05). XK is supported by the Bairen program of the Chinese Academy of Sciences, the foundation for the author of CAS excellent doctoral dissertation, and NSFC project (No. 11073055). GL is supported by the Bairen program of the Chinese Academy of Sciences.

\clearpage

\begin{figure}
\epsscale{0.80}
\includegraphics[width=1\textwidth]{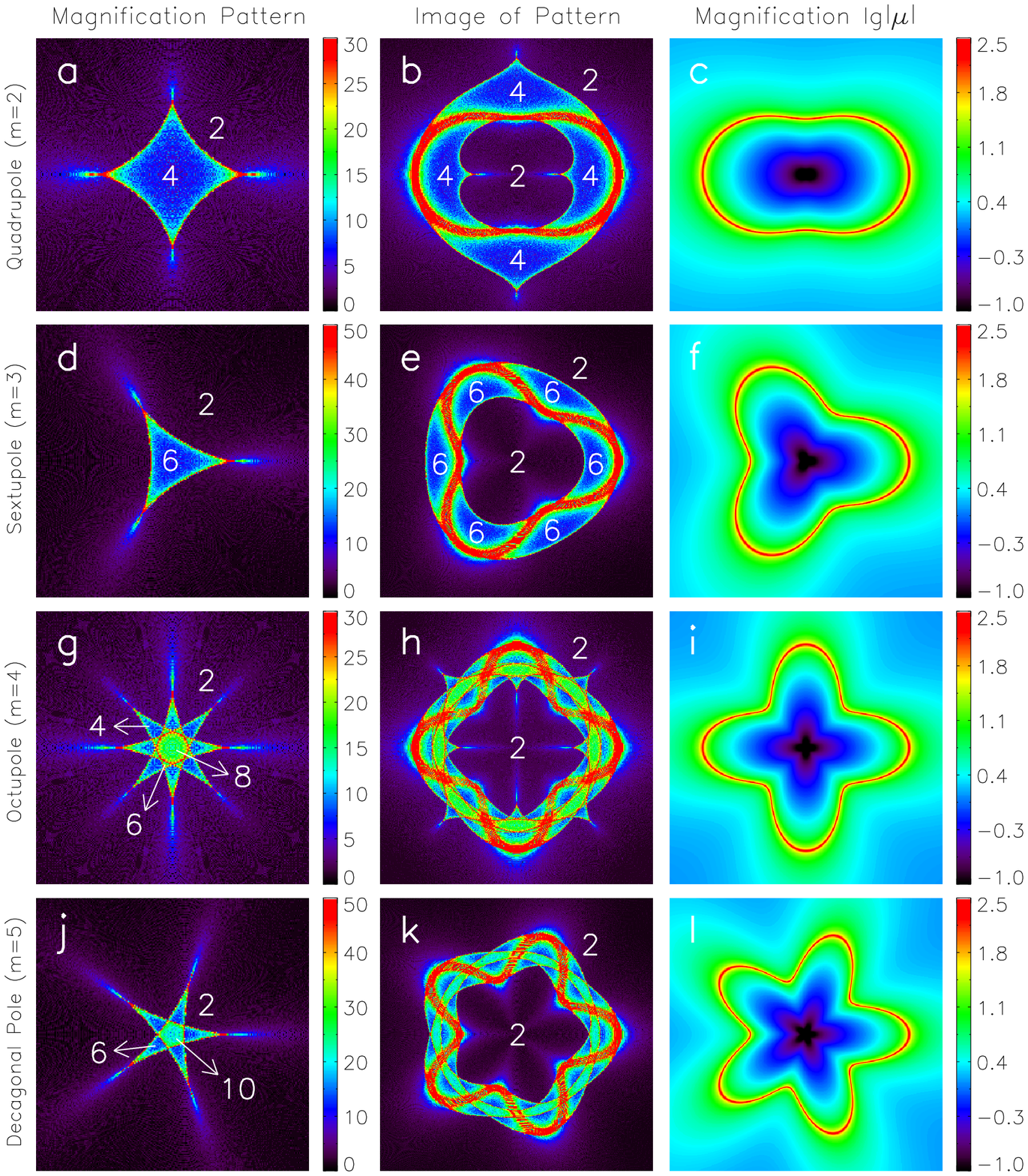}
\caption{Magnification patterns (left column) and their images (middle column) as well as the magnification maps (right column) for $k_m=0.3$, where $m=2$, 3, 4 and 5 are corresponding to the rows displayed from top to bottom respectively. The numbers labled on the panels denote the regions of different image multiplicities, and the color-bars indicate the magnification level.\label{fig1}}
\end{figure}

\begin{figure}
\epsscale{0.80}
\includegraphics[width=1\textwidth]{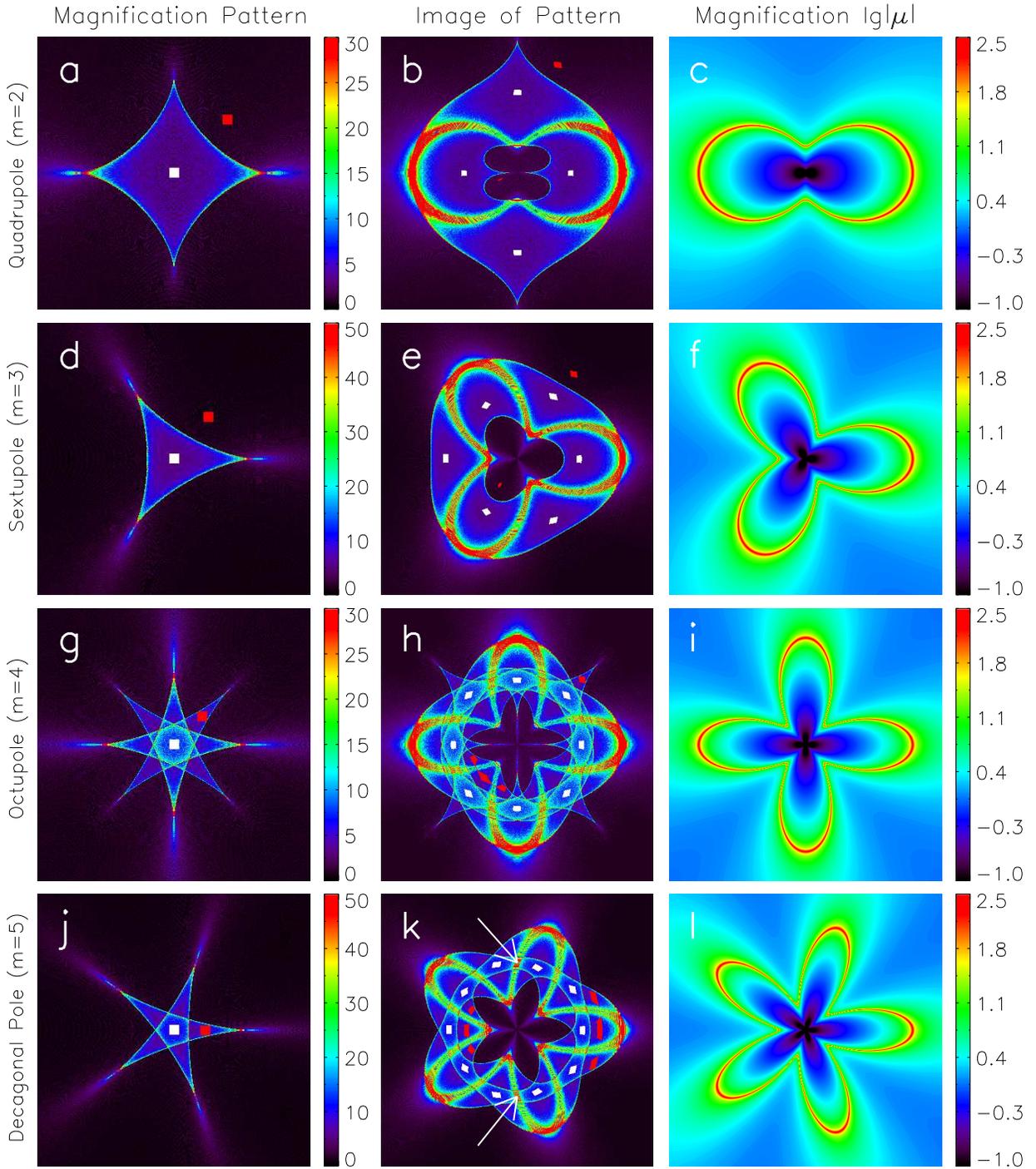}
\caption{Similar to Figure 1, but for $k_m=0.6$ in contrast. The white and red areas on the middle column are the images of the white and red squares on the left column.\label{fig2}}
\end{figure}

\begin{figure}
\epsscale{0.80}
\includegraphics[width=1\textwidth]{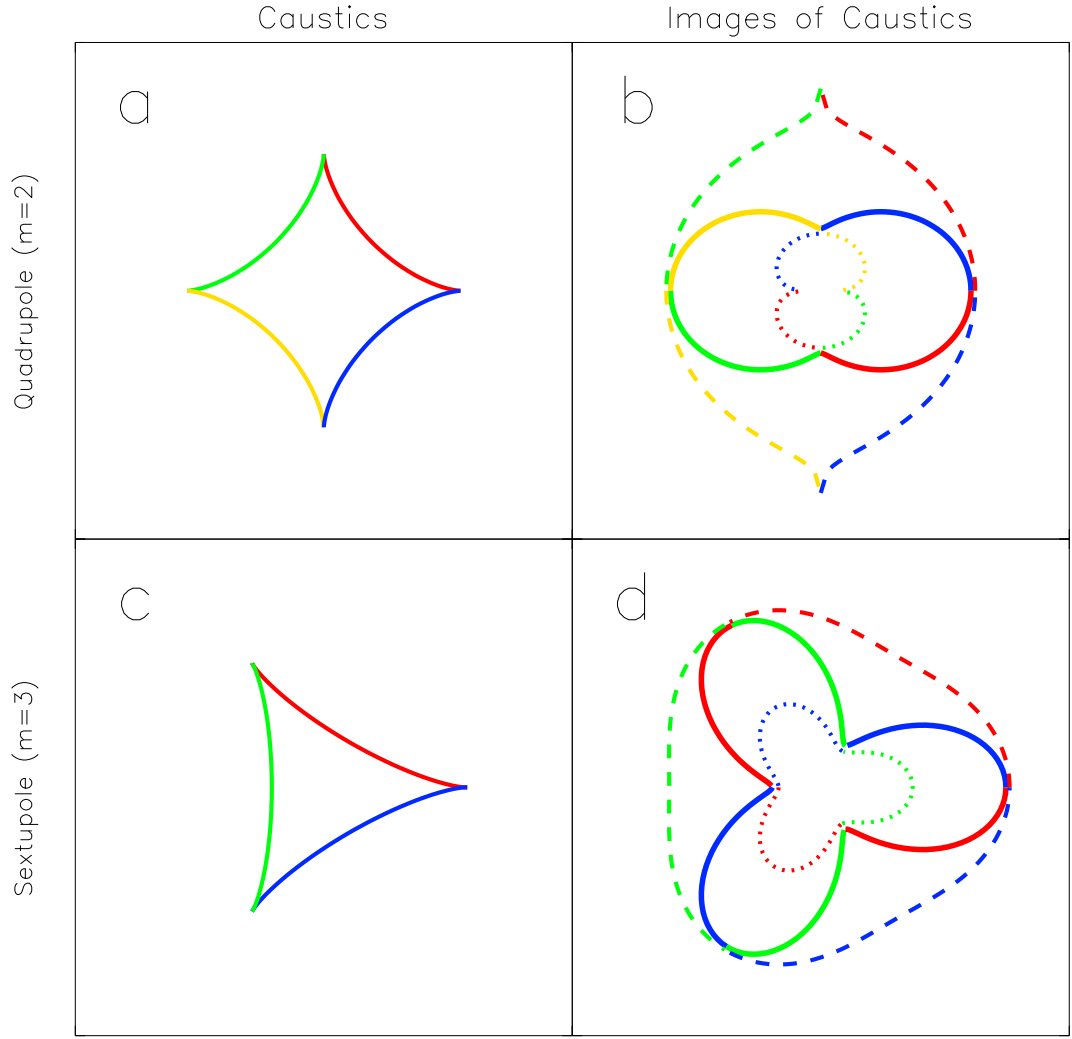}
\caption{The tangential caustics of $m=2$ and $m=3$ mode lenses (left column) and their images (right column). In the right column, the thick solid curve denote the critical curve, while the dashed and doted curves denote the outer and inner transition loci respectively.\label{fig3}}
\end{figure}

\begin{figure}
\epsscale{1.00}
\includegraphics[width=1\textwidth]{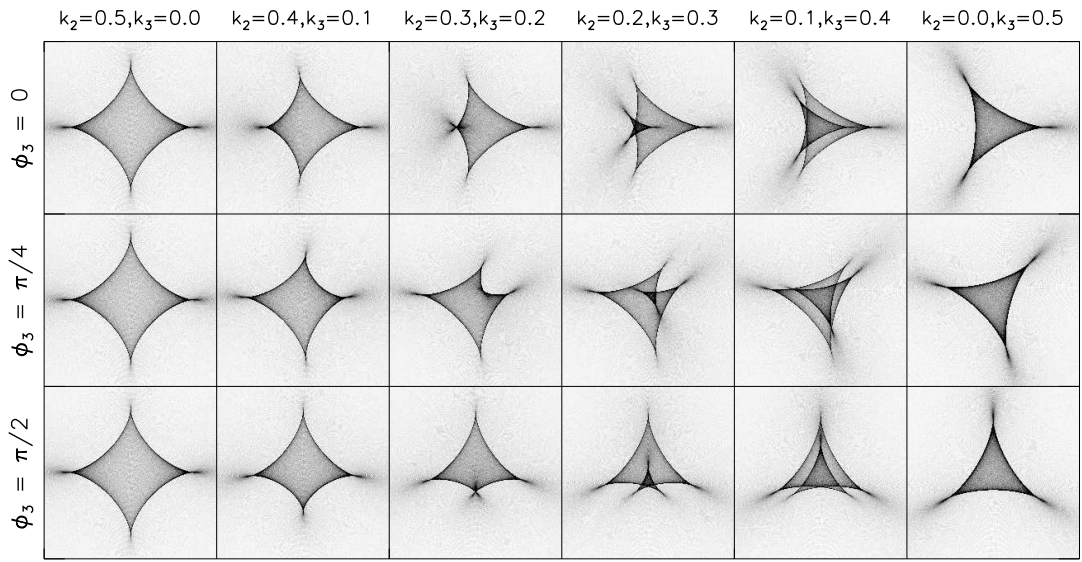}
\caption{The magnification patterns of $m=2$ lenses perturbed by the $m=3$ components. From top to bottom the perturbations phases is 0, $\pi/4$, $\pi/2$ respectively, and from left to right $k_3$ is set from 0.0 to 0.5 respectively, with constraint of $k_2 + k_3 =0.5$.\label{fig4}}
\end{figure}

\begin{figure}
\epsscale{1.00}
\includegraphics[width=1\textwidth]{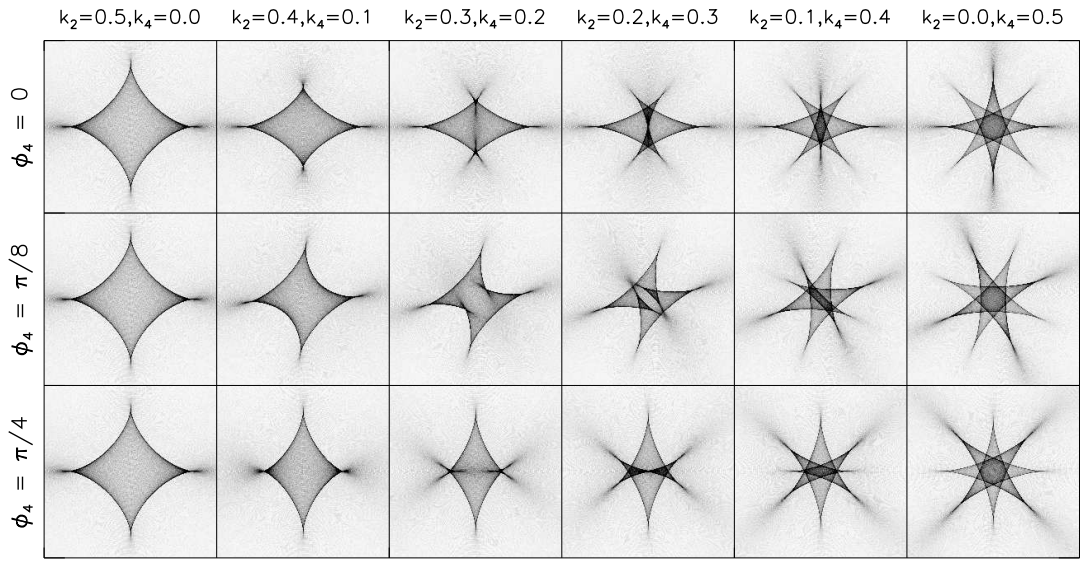}
\caption{The same as Figure 4, except perturbed by the $m=4$ components, and the perturbations corresponding to the phases 0, $\pi/8$, $\pi/4$ respectively.\label{fig5}}
\end{figure}

\begin{figure}
\epsscale{1.00}
\includegraphics[width=1\textwidth]{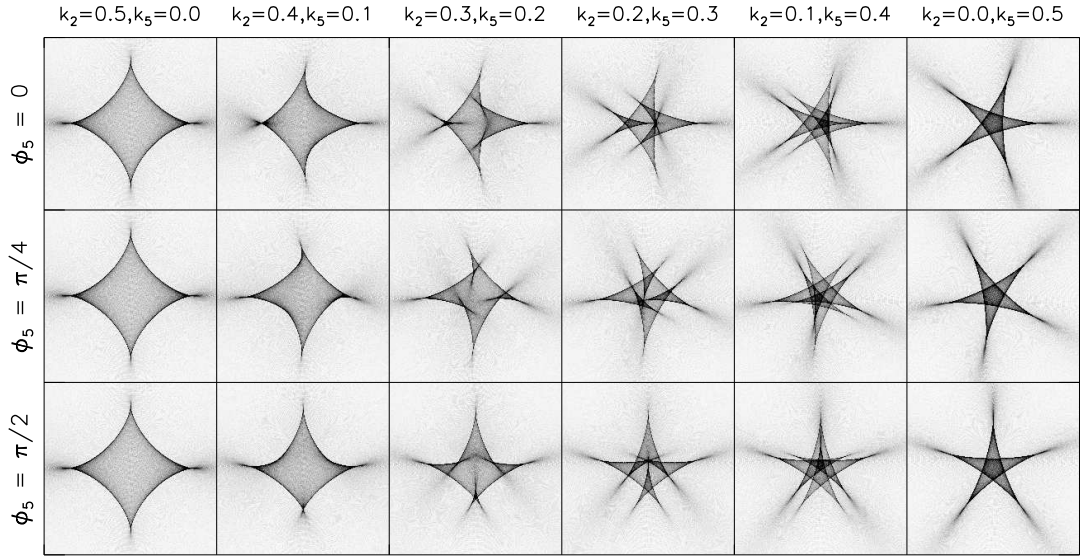}
\caption{The same as Figure 4, except perturbed by the $m=5$ components.\label{fig6}}
\end{figure}




\clearpage


\end{document}